\documentstyle[12pt]{article}
\begin{document}
\hfill$DIAS-STP-94-18$

\hfill {Revised \vspace{20mm} version}

\centerline{\LARGE Yang-Baxterization of the
$B{\cal H}$
\vspace{10mm}algebra.}
\centerline{\large G.A.T.F.da Costa}
\centerline{\large Dublin Institute for Advanced Studies}
\centerline{\large 10 Burlington Road,Dublin 4,Ireland.}
\centerline{\large May 14,\vspace{10mm}1994}

\begin{abstract}
The $B{\cal H}_{n}(l,m)$
algebra
is defined by two sets of generators one of
which satisfy
the relations of the braid group and the other the
relations of the Hecke algebra of projectors.These
algebras are then combined by additional relations
in a way which generalize the Birman-Wenzl algebra
${\cal C}_{n}(l,m)$.
In this paper we Yang-Baxterize the algebra $B{\cal H}$
and
compute solutions of the Yang-Baxter equation.The
solutions found are expressed algebraically in terms
of the generators of the algebra.The expression
generalizes the
known one for the
algebra ${\cal C}_{n}(l,m)$.
\end{abstract}

\section{Introduction}

The Yang-Baxter equation (YBE)
$$
\check{R}_{1}(x)\check{R}_{2}(xy)\check{R}_{1}
=\check{R}_{2}(y)\check{R}_{1}(xy)\check{R}_{2}(x)
\eqno(1)
$$
plays a central role as the main criterium for
integrability of lattice models
in statistical mechanics.
The equation bares a resemblance to the
generating relations of the Artin's
braid group algebra $B_{n}$ defined by the
generators $1,\sigma_{i},i=1,...,n-1$ and
relations
$$
{\sigma}_i{\sigma}_{i+1}{\sigma}_i
={\sigma}_{i+1}{\sigma}_{i}{\sigma}_{i+1},
$$
$$
{\sigma}_i{\sigma}_j={\sigma}_j{\sigma}_{i},\hspace{2mm}
{\mid}{i-j}{\mid}\geq2,
\eqno(2)
$$
the difference being that YBE depends explicitly
on a spectral parameter.However,
many examples of solutions
are known which in the limit when the
parameter goes to zero or infinity the matrices
satisfying the equation become a representation
of the braid group.
In [1-2] Jones considered the inverse problem
of finding
solutions of the
YBE equation
from representations of the braid group.
He named this procedure of constructing
solutions of YBE,Yang-Baxterization.

The purpose of this note is to discuss the
Yang-Baxterization of the algebra $B{\cal H}_{n}(l,m)$.This
algebra first appeared on [5] in association
with a class of lattice models (see section 3) and
it has important mathematical properties such
as connections with topological invariants of knots
and graph embeddings [5].
It
depends on two parameters l and m
and it is defined by the generators $1,\sigma_{i},
H_{i},i=1,...,n-1$ satisfying the relations of
the braid algebra (2) and of the Hecke algebra ${\cal H}(q)$
$$
H_{i}H_{i\pm1}H_{i}-H_{i}
=H_{i\pm1}H_{i}H_{i\pm1}-H_{i\pm1}
\eqno(3.1)
$$
$$
H^{2}_{i}=qH_{i}
\eqno(3.2)
$$
$$
H_{i}H_{j}=H_{j}H_{i},\hspace{2mm}
{\mid}{i-j}{\mid}\geq2,
\eqno(3.3)
$$
with $q=1 + m^{-1}(l-l^{-1})$ and the additional relations
$$
\sigma_{i}-\sigma^{-1}_{i}=m(1-H_{i})
\eqno(3.4)
$$
$$
\sigma_{i\pm1}\sigma_{i}H_{i\pm1}
=H_{i}\sigma_{i\pm1}\sigma_{i}
\eqno(3.5)
$$
$$
\sigma_{i\pm1}\sigma_{i}H_{i\pm1}-H_{i}H_{i\pm1}
=\sigma_{i}\sigma_{i\pm1}H_{i}-H_{i\pm1}H_{i}
\eqno(3.6)
$$
$$
\sigma_{i\pm1}H_{i}\sigma_{i\pm1}
-\sigma^{-1}_{i}H_{i\pm1}\sigma^{-1}_{i}
=\sigma_{i}H_{i\pm1}\sigma_{i}
-\sigma^{-1}_{i\pm1}H_{i}\sigma^{-1}_{i\pm1}
\eqno(3.7)
$$
$$
\sigma_{i\pm1}H_{i}H_{i\pm1}-\sigma^{-1}_{i}H_{i\pm1}
=\sigma_{i}H_{i\pm1}H_{i}-\sigma^{-1}_{i\pm1}H_{i}
\eqno(3.8)
$$
$$
H_{i\pm1}H_{i}\sigma_{i\pm1}-H_{i\pm1}\sigma^{-1}_{i}
=H_{i}H_{i\pm1}\sigma_{i}-H_{i}\sigma^{-1}_{i\pm1}
\eqno(3.9)
$$
$$
\sigma_{i}H_{i}=H_{i}\sigma_{i}=\frac{1}{l}H_{i}
\eqno(3.10)
$$
$$
H_{i}\sigma_{i\pm1}H_{i}-lH_{i}
=H_{i\pm1}\sigma_{i}H_{i\pm1}-lH_{i\pm1}
\eqno(3.11)
$$
$$
\sigma^{2}_{i}=m(\sigma_{i}-\frac{1}{l}H_{i})+1
\eqno(3.12)
$$
$$
\sigma^{3}_{i}=(m+\frac{1}{l})\sigma^{2}_{i}
-(1-\frac{m}{l})\sigma_{i}
-\frac{1}{l}
\eqno(3.13)
$$
The $B{\cal H}$ algebra is closely associated
with the Birman-Wenzl algebra ${\cal C}_{n}$ [3]
whose Yang-Baxterization was first studied
by Jones in [1,2] and then in [4].
Indeed,${\cal C}_{n} \subset
B{\cal H}$ and its relations can be obtained
from relations (3) above by taking
$H_{i} \rightarrow E_{i}$
and both sides of relations (3.1),(3.6-9) and (3.11)
equal to zero.
Relations (3) are not all independent
but they are given here
because they are used in section (2) in
the derivation of
our main result:solutions of equation (1)
which have support in the algebra.These are
solutions of the trigonometric type
and are given in algebraic form,there is,
in terms of the generators $\sigma_{i},H_{i}$
of the algebra (2-3).
In section (3) we discuss an explicit realization.

\section{Main result}

Let $\sigma_{i}$ be a solution of
(2) satisfying the cubic relation
with three distinct eigenvalues $\lambda_{1},\lambda_{2}$
and $\lambda_{3}$
$$
(\sigma_{i}-\lambda_{1})(\sigma_{i}-\lambda_{2})
(\sigma_{i}-\lambda_{3})
=0
\eqno(4)
$$
The following result was proved in [4]:

{\bf Theorem 1:}If $\sigma_{i}$ satisfy
(4) and the identity
$$
f_{3}^{+} \theta_{3}^{+} + f_{3}^{-} \theta_{3}^{-}
+ f_{2} \theta_{2} +f_{1}^{+} \theta_{1}^{+}
+ f_{1}^{-} \theta_{1}^{-}=0
\eqno(5.1)
$$
where
$$
\theta_{3}^{\pm}=\sigma_{1}^{\pm}\sigma_{2}^{\mp}\sigma_{1}^{\pm}
-\sigma_{2}^{\pm}\sigma_{1}^{\mp}\sigma_{2}^{\pm}
$$
$$
\theta_{2}=\sigma_{1}\sigma_{2}^{-1}-\sigma_{2}\sigma_{1}^{-1}
+\sigma_{2}^{-1}\sigma_{1}-\sigma_{1}^{-1}\sigma_{2}
$$
$$
\theta_{1}^{\pm}=\sigma_{1}^{\pm}-\sigma_{2}^{\pm}
\eqno(5.2)
$$
and $f_{3}^{\pm},f_{2},f_{1}^{\pm}$ are given by
$$
f_{3}^{+}=\frac{\lambda_{1}}{\lambda_{3}^{2}},
f_{3}^{-}=-\frac{\lambda_{1}^{2}}{\lambda_{3}}
$$
$$
f_{2}=-\frac{\lambda_{1}}{\lambda_{3}}
(1 + \frac{\lambda_{1}}{\lambda_{2}}
+ \frac{\lambda_{2}}{\lambda_{3}}
+ \frac{\lambda_{1}}{\lambda_{3}})
$$
$$
f_{1}^{\pm}=\mp \lambda_{2}^{\mp} f_{2}
\eqno(5.3)
$$
then,
$$
\check{R}_{i}(x)=A(x) \sigma_{i} + B(x) I + C(x) \sigma_{i}^{-1}
\eqno(6)
$$
is a solution of the YBE
satisfying the following three conditions:

1)the boundary condition :
$$
\check{R}_{i}(0) \sim \sigma_{i}
\eqno(7.1)
$$

2)the initial condition:
$$
\check{R}_{i}(1) \sim I
\eqno(7.2)
$$

3)the unitarity condition:
$$
\check{R}_{i}(x)\check{R}_{i}(x^{-1})=f(x) I,
\eqno(7.3)
$$
for some function f(x)
and
the coefficients A,B,and C
can be chosen of the following form:
$$
A(x)= -\lambda_{3}^{-1}(x-1),C(x)=\lambda_{1}x(x-1)
$$
$$
B(x)=(1+ \frac{\lambda_{1}}{\lambda_{2}}+
\frac{\lambda_{1}}{\lambda_{3}}+ \frac{\lambda_{2}}{\lambda_{3}})x
\eqno(8.1)
$$
$$
A(x)= -\lambda_{3}^{-1}(x-1),C(x)=\lambda_{2}x(x-1)
$$
$$
B(x)=(1+ \frac{\lambda_{2}}{\lambda_{1}}
+\frac{\lambda_{2}}{\lambda_{3}}+ \frac{\lambda_{1}}{\lambda_{3}})x
\eqno(8.2)
$$
%$$
%A(x)= -\lambda_{2}^{-1}(x-1),C(x)=\lambda_{1}x(x-1)
%$$
%$$
%B(x)=(1+ \frac{\lambda_{1}}{\lambda_{3}}
%+\frac{\lambda_{1}}{\lambda_{2}}+ \frac{\lambda_{3}}{\lambda_{2}})x
%\eqno(8.3)
%$$

It was also proved in [4] that a sufficient condition
for a solution $\sigma_{i}$ of (4) to satisfy identity
(5.1) is that it admits the algebraic structure
of the Birman-Wenzl algebra ${\cal C}_{n}$.Our
main result consists in the proof that (5.1)
is still satisfied if $\sigma_{i}$ admits the
structure of the more general $B{\cal H}$ algebra.This
is stated in the following theorem:

{\bf Theorem 2}:Suppose $\sigma_{i}$ and $H_{i}$
close the relations (2) and (3) of algebra $B{\cal H}(l,m)$ with
$l=\lambda_{3}^{-1}$,$m=\lambda_{1}+ \lambda_{2}$ and
$\lambda_{1} \lambda_{2}=1$.Then,identity (5.1) is satisfied
and $\check{R}_{i}$ given by (6-8) is a solution
of YBE (1) with support in $B{\cal H}$.

{\bf Proof:}

Using solely the relations (2) and (3)
of the algebra relations
(5.2) can be rewritten as follows:
$$
\theta_{3}^{+}=m^{2}\{\sigma_{2}-\frac{1}{l}H_{2}
-\sigma_{1} +\frac{1}{l} H_{1}
-\sigma_{2}H_{1}
$$
$$
+ \sigma_{1}H_{2} - H_{1}\sigma_{2}
+mH_{1}+H_{2}\sigma_{1}-mH_{2}\}
\eqno(9.1)
$$
$$
\theta_{3}^{-}=m^{2}\{\sigma_{2}-\sigma_{2}H_{1}
-H_{1}\sigma_{2}+lH_{1}
$$
$$
-lH_{2}-\sigma_{1}+\sigma_{1}H_{2} + H_{2}\sigma_{1}\}
\eqno(9.2)
$$
$$
\theta_{2}=m\{-2\sigma_{1}+2\sigma_{2}+\sigma_{1}H_{2}
-\sigma_{2}H_{1}
$$
$$
+H_{2}\sigma_{1}
-H_{1}\sigma_{2}\}
\eqno(9.3)
$$
$$
\theta_{1}^{+}=\sigma_{1}-\sigma_{2}
\eqno(9.4)
$$
$$
\theta_{1}^{-}=\sigma_{1}-\sigma_{2}+mH_{1}-mH_{2}
\eqno(9.5)
$$
Upon substitution of these relations in the
left hand side of (5.1) it is observed
that all terms cancel out and the
identity is satisfied.Moreover,$\check{R}_{i}$ is
a solution of (1) with A,B,C given by (8).

In terms of the generators $\sigma_{i}$ and
$H_{i}$ of the algebra $B{\cal H}$,the
solutions (6) can  be expressed as
$$
\check{R}_{i}(x)=(A(x)+C(x))\sigma_{i}
+ (B(x)-mC(x))I + mC(x)H_{i}
\eqno(10)
$$
for A,B,C given by (8.1) and (8.2).
Thus,two solutions of YBE are associated to
every representation of the braid group
satisfying the $B{\cal H}$ algebra.
Since every representation of the
Birman-Wenzl algebra
${\cal C}_{n}$
is also a representation of $B{\cal H}$ the
solutions found in [4] are automatically taken into
account by
(10).

\section{An example}

A wide class of lattice models have a partition
function which can be expressed in general form
as
$$
{\cal Z}=Tr\{\prod _{j=1}^{N-1}(D+A\theta_{2j})
\prod _{j=1}^{N}(C+B \theta_{2j-1})\}^{M}
\eqno(11)
$$
where the letters are parameters associated with
the anisotropies of the model and M and N
are the numbers of rows and columns of a square
lattice.The $\theta's$ are given either by a
representation of the Temperley-Lieb or of Hecke algebras.
Let's consider the case when $D=C=t$ and $A=B=t^{-1}$
and take $\theta_{i}=H_{i} \in {\cal H}_{n}$
satisfying relations (3.1-3).

A
simple check shows that the $H_{i}$ close with the
$$
\sigma_{i}^{\pm}=t^{\pm} + t^{\mp}H_{i}
\eqno(12)
$$
relations (2-3)
with $m=t - t^{-1}$ and $l=-t^{3}$ and $q=-(t^{2}+t^{-2})$.
The literature on lattice models provide plenty examples
of representations for the generators $H_{i}$ and from them
using (12) we can obtain representations for the algebra
$B{\cal H}_{n}$.For instance [6],
$$
H_{i}= \sum_{k=1}^{Q} \sum_{l=1}^{Q}
I^{(1)}  \otimes ... \otimes (x^{\epsilon(k,l)/2}
E_{kk}^{(i)}  \otimes  E_{ll}^{(i+1)}
+ E_{kl}^{(i)} \otimes E_{lk}^{(i+1)})
 \otimes ... I^{(n)}
\eqno(13)
$$
where the sums over l and k are such that
$l \neq k,i=1,2,...,n-1$,$E_{ab}$ is the matrix
whose elements are $(E_{ab})_{cd}=\delta_{ac}
\delta_{bd}$.The number Q is arbitrary and
the superscripts label the slots where
the factors are in the tensor product and
$\epsilon(k,l)=-1$ if $k < l$ and
$+1$ if $k > l $.The relation of ${x}$ with ${t}$ is
given by $-t^{\pm 2}= x^{\mp 1/2}$.

It is remarkable that the $\sigma_{i}$ given by (12)
admit the cubic relation
(3.13) or (4) with eigenvalues given by
$\lambda_{1}=t,\lambda_{2}=-t^{-1},\lambda_{3}=
-t^{3}$.
They can be
Yang-Baxterized to produce
new solutions of YBE (1) using (8.1-2).
For A,B,C given by (8.1) and using (10)
we get
$$
\check{R}_{i}(x)=(x-t^{2})(x+\frac{1}{t^{4}})
+ (x-1)(xt^{2} + \frac{1}{t^{4}})H_{i}
\eqno(14)
$$
and,for A,B,C given by (8.2),
$$
\check{R}_{i}(x)=(1+ \frac{1}{t^{4}})x-\frac{1}{t^{2}}(1+x^{2})
+ (\frac{1}{t^{4}}-x)(x-1)H_{i}
\eqno(15)
$$
The elements $g_{i}=t.\sigma_{i}$ where
$\sigma_{i}$ are as in (12) admit the relations
of the braided Hecke algebra,that is,
the $g_{i}$ satisfy relations (2) and the quadratic
relation $g_{i}^{2}=(t^{2}-t^{-2}) g_{i} + 1$.
However,the solution of YBE obtained by the Baxterization
of this algebra (see [1])
with $g_{i} =t\sigma_{i}$
can not be mapped consistently on the solutions
(14) or (15) or vive-versa.

\end{document}